\documentstyle[prcrc,11pt,fleqn]{article}

\def\gsim{\mbox{\raisebox{-1.0ex}{$\stackrel{\textstyle >}{\textstyle \sim}$
}}}

\newcommand{\phidot}{\dot{\phi}}
\newcommand{\PSbox}[3]{\mbox{\rule{0in}{#3}\includegraphics{#1}\hspace{#2}}}

\title{Comments on non-Gaussian density perturbations and 
   the production of primordial black holes}

\author{James S. Bullock, Joel R. Primack \address{Physics Department, 
        University of California Santa Cruz, \\ 
        Santa Cruz, CA 95064}\thanks{We thank Pavel Ivanov for
        thoughtful comments and discussion.}}

\begin{document}
% typeset front matter
\maketitle

\begin{abstract}
We review the basic arguments for the likelihood of
non-Gaussian density perturbations in inflation models with
primordial black hole (PBH) production.  We 
discuss our derived distributions of field fluctuations
and their implications, specifically commenting on
the fine-tuning problem.
We also discuss how the derived distributions may be affected
when linked to metric perturbations.  
While linking the metric perturbations
to field fluctuations in a nonlinear way may be important for determining
exact probability distributions, the correct
mapping is not self-evident.  The calculation of P. Ivanov, which yields
a skew positive distribution, is based on
an \textit{ansatz} for the behavior of the nonlinear metric perturbation.
We note that the ``natural'' generalization of the gauge-invariant 
formalism favored by Bond and Salopek 
yields an effective linear link between the distribution of field fluctuations
and metric perturbations during inflation.

\end{abstract}

\section{Primordial black holes and non-Gaussianity}
Any (realistic) PBH-producing model of inflation must yield a power spectrum
that is consistent with COBE on large scales ($\delta = \delta \rho /
\rho \sim 10^{-5}$),
while at the same time giving $\delta \sim 0.01$ on some small scale.
Such models are possible, however, as we explain below and 
in detail~\cite{us}, 
the large-amplitude small-scale power can lead to non-Gaussian fluctuation
statistics.

Let $\phi$ represent the inflaton.   During inflation,
quantum fluctuations in the inflaton, $\delta \phi_{Q}\sim H$, 
generate density perturbations. 
The density fluctuations at horizon crossing map to the following 
expression as the field rolls down the potential: 
$\delta \sim H^2/\phidot$, where $\phidot$ is the time derivative
of $\phi$.  If the classical trajectory of $\phi$
is unaffected by quantum fluctuations, the fluctuations 
evolve linearly and Gaussian statistics prevail.  This is true
in most models of inflation.  
Over a Hubble time, the change in $\phi$ due to its classical 
trajectory is $\Delta\phi_{CL} \sim \phidot H^{-1}$.
Gaussian statistics are then likely provided
$\delta \phi_{Q} / \Delta \phi_{CL} \ll 1$. But 
notice the following relation: $\delta 
\sim H^2 / \phidot \sim \delta \phi_{\rm Q} / \Delta \phi_{\rm CL}.$
The Gaussian criterion breaks down as the amplitude of $\delta$
gets large.  Therefore, if we do have some inflation
model that produces the $\delta \sim 0.01$ needed for PBH formation,
we may expect non-Gaussian statistics to be important.

\begin{figure}[htb]
\begin{minipage}[t]{75mm}
\PSbox{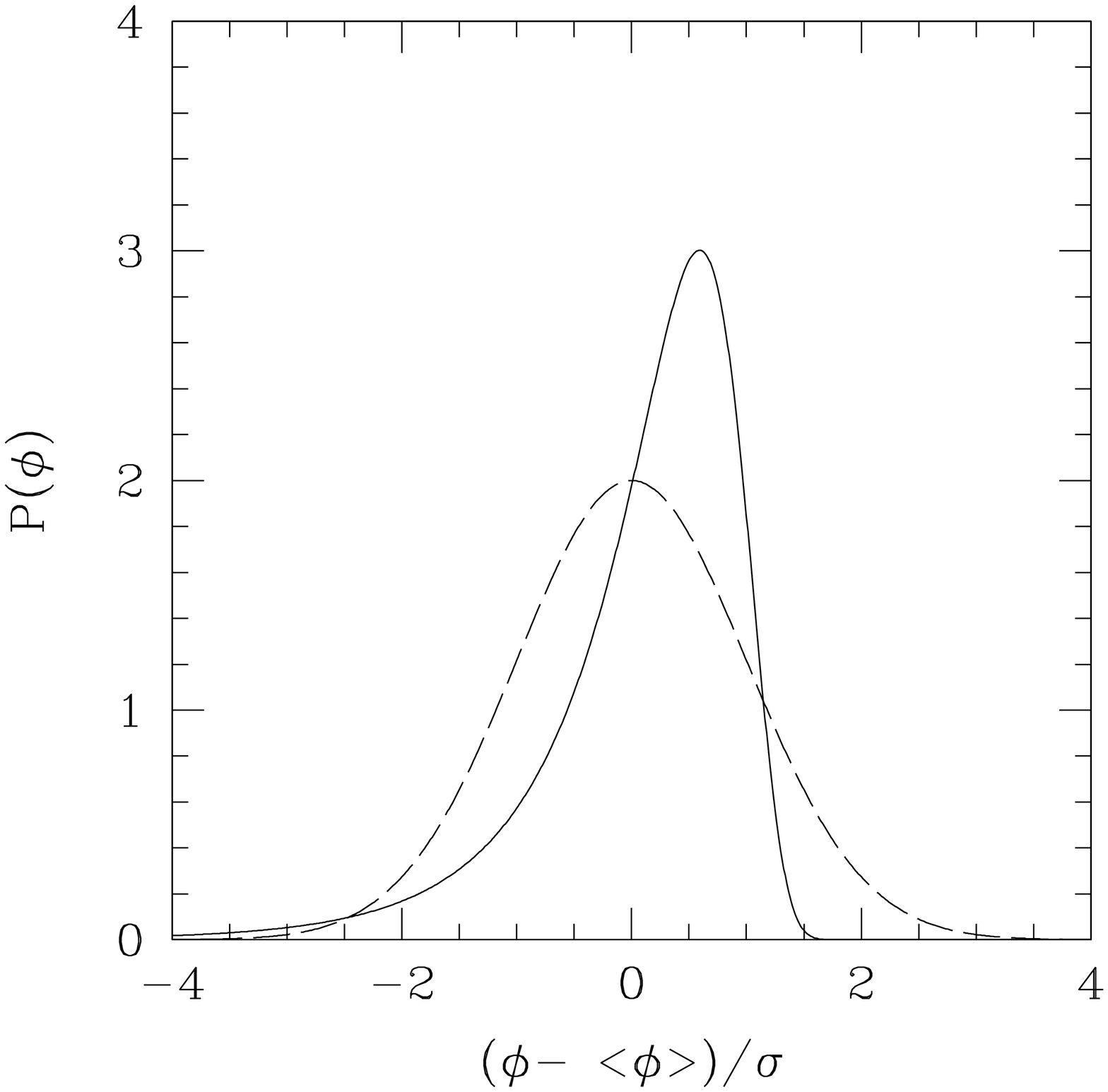 hoffset=-10 voffset=-60 hscale=35 vscale = 35}{74mm}{74mm}
\caption{Our calculated distribution of field fluctuations (solid line) 
compared to a Gaussian 
with the same mean and standard deviation (dashed line).}
\label{fig:dist}
\end{minipage}
\hspace{\fill}
\begin{minipage}[t]{75mm}
\PSbox{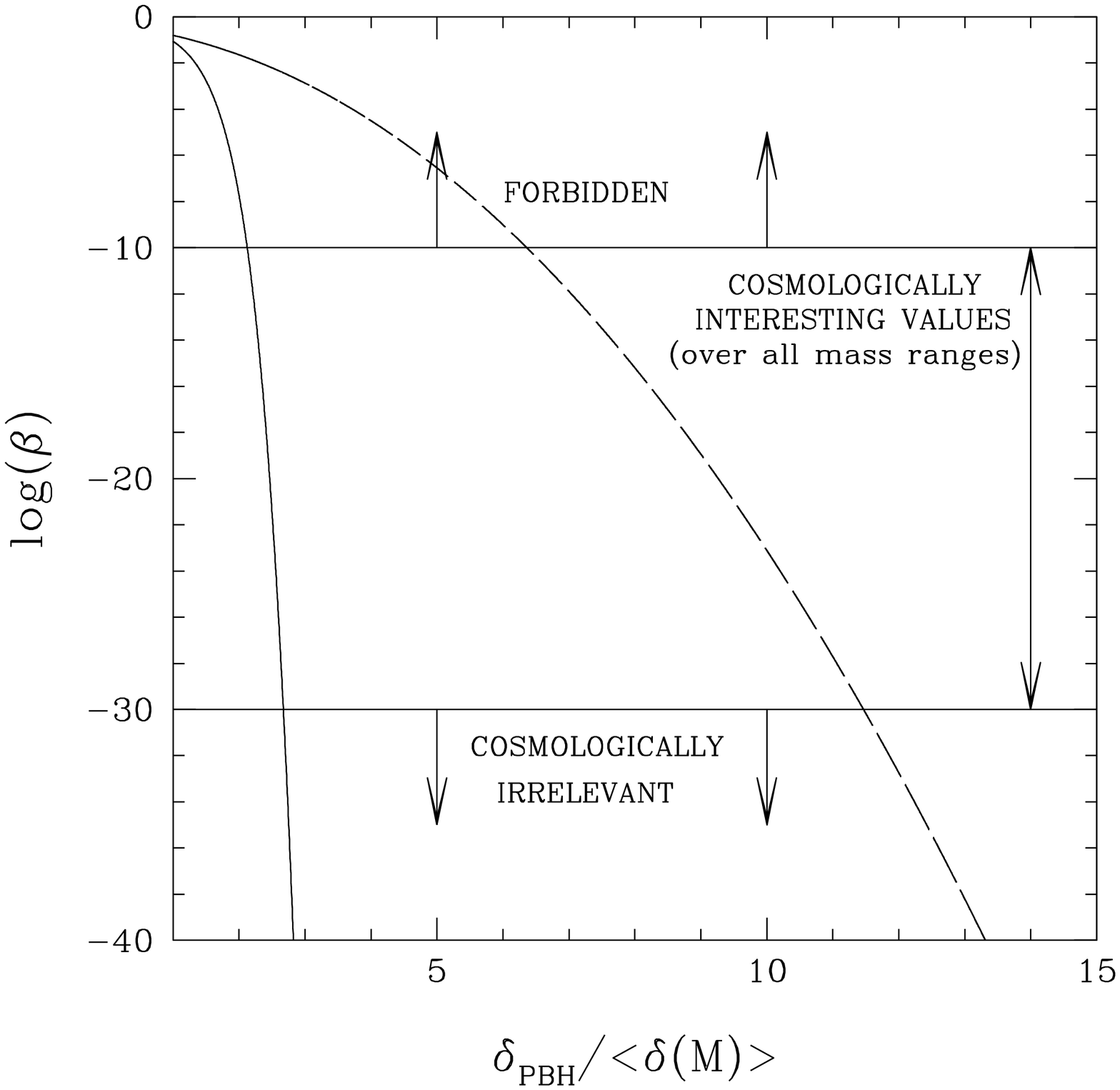 hoffset=-10 voffset=-60 hscale=35 vscale = 35}{74mm}{74mm}
\caption{The initial mass fraction of PBHs produced, $\beta$,  versus
standard deviation fluctuations needed for PBH formation.
The dashed line is derived from a Gaussian distribution and the 
solid line results from the non-Gaussian distribution
shown in Figure 1.  
}
\label{fig:beta}
\end{minipage}
\end{figure}

\section{Toy model results}
In order to test the the qualitative ideas presented in the
previous section, we have constructed several (toy) PBH-producing
inflation models and used stochastic inflation calculations
to find the associated  fluctuation statistics~\cite{us}.  
For example, Figure 1 shows a typical probability distribution 
of fluctuations from one of our models together with a Gaussian
distribution with the same standard deviation.
Note that the Gaussian over produces large
fluctuations relative to the derived distribution.  See 
Ref.~\cite{ambleside} for a review.

One important consequence of this non-Gaussian result is that it
intensifies the fine-tuning problem in PBH formation.  The fine-tuning
problem arises because the height of the peak in power, $\delta(M)$,
(where $M$ is the mass of PBHs produced) 
must be fine-tuned
to an extremely precise value in order to obtain a cosmologically 
relevant fraction of PBHs without over-production.

Limits on PBH abundances are often quoted in terms of the initial
mass fraction of PBHs $\beta$~\cite{us}. Primordial black
holes form when 
$\delta \gsim \delta_{\rm PBH} \sim 1/3$.  Therefore, if 
$P(\delta)$ is the distribution of 
fluctuations on a mass scale 
$M$ with an rms deviation $\delta(M)$,
the value of $\beta$ is determined by integrating over the
high-$\delta$ tail of the distribution, ($\delta \gsim \delta_{\rm PBH}$).  
The result will depend on the ratio, $\delta_{PBH} / \delta(M)$ (
\textit{i.e.} the size of a PBH-producing fluctuation relative to the rms 
deviation).

Figure 2 shows $\beta$ plotted against this ratio 
for a Gaussian distribution (dashed line) and the non-Gaussian case
(solid line) plotted in Figure 1. 
The horizontal lines show the range of
cosmologically interesting values of $\beta$ over all
relevant PBH masses.  We see that the value of $\beta$ changes
drastically for small changes in the relevant ratio for both
distributions, but is even worse for the non-Gaussian expectation.

For the non-Gaussian case, $\delta_{PBH} / \delta(M)$
must fall between 2 and 3 in order to be cosmologically
relevant at any mass scale.  This 
precision must occur within a spike
in power that is $\sim 4$ orders of magnitude above the COBE normalization.
The upshot is that any inflation model that attempts
to construct cosmologically relevant PBH formation must be tuned
to at least $10^{-4}$, and if we restrict ourselves to any specific
mass scale, the required tuning becomes much worse. 

\section{Nonlinear metric perturbations}

In the standard linear treatment of inflationary fluctuations
$\delta \phi \propto \delta$.  So, to first order
we may infer the probability distribution of 
density fluctuations by deriving the distribution of $\phi$ fluctuations.
However, as mentioned in Ref.~\cite{us}, since the density fluctuations are large,
any nonlinearity in the connection between metric perturbations
and field fluctuations may change the distribution
shape.  One should formally understand this relationship by
coupling field fluctuations to the Einstein equations during inflation.

This problem was explicitly noted by Ivanov~\cite{ivanov}, who presented
a PBH-producing model using a flat plateau feature in the
inflaton potential.  Because the region is exactly flat, the Ivanov
model yields a near Gaussian 
distribution of field fluctuations or, equivalently,
a near Gaussian distribution in the stochastic time, $\delta t$, 
that the inflaton spends in the flat region of the
potential (the region associated with PBH production).

Ivanov chooses to relate metric perturbations to this stochastic 
time delay using
the \textit{ansatz}: $h = \exp(H \delta t) - 1$, where
$H$ is the value of the Hubble parameter associated with the
flat region of the potential.
In the limit of small $h$, 
this relation reduces to the standard gauge invariant quantity.  
The proposed exponential relation between $\delta t$ and $h$, when 
combined with
the near Gaussian distribution in $\delta t$,
yields a positively skewed distribution of $h$.
Note, however, that this result is based entirely
on the \textit{ansatz}.  The relation favored by Bond and Salopek~\cite{bond}
is $h = H \delta t$, which produces a linear transformation and,
thus, an effectively Gaussian distribution in $h$.

Understanding the nature of the mapping between derived
$\delta \phi$ statistics and $\delta$ statistics could
be important in determining the exact nature of fluctuation 
distributions associated with PBHs.  
However, since the standard treatment yields a linear
relationship, the exponential relation discussed above should, perhaps,
be viewed as an extreme possibility.
We are currently attempting to clarify this issue, but advocate the linear
mapping as a first approximation until a more strongly  motivated relation
can be found.

\end{document}